\documentclass[journal]{IEEEtran}      
\ifCLASSINFOpdf

\else

\fi

\usepackage[cmex10]{amsmath}
\usepackage{amssymb,wrapfig,multirow,amsfonts,amsthm,mathrsfs}
\usepackage{multicol}
\usepackage{epstopdf}\usepackage{footnote}
\usepackage{tikz}
\usepackage{filecontents,lipsum,graphicx,xcolor}

\begin{document}
\title{Quantum-Secure Certificate-Less Conditional Privacy-Preserving Authentication for VANET}

\author{Girraj~Kumar~Verma, Nahida~Majeed~Wani, and Prosanta~Gope, \IEEEmembership{Senior Member,~IEEE} 
\thanks{G. K. Verma and N. M. Wani are with the Department of Mathematics, Amity School of Engineering and Technology, Amity University, Madhya Pradesh, Gwalior, India (Email:gkverma@gwa.amity.edu, girrajv@gmail.com, nahidaoct@gmail.com)}
\thanks{P. Gope is with the Department of Computer Science, University of Sheffield
Sheffield S1 4DP, United Kingdom (Email: p.gope@sheffield.ac.uk).}
}

\maketitle                   
\begin {abstract}
Vehicular Ad-hoc Networks (VANETs) marked a pronounced change in the Intelligent Transport System and Smart Cities through seamless vehicle communication to intensify safety and efficacy. However, a few authentication schemes have been devised in the literature to ensure the authenticity of the source and information in the post-quantum era. The most popular base for such construction is lattice-based cryptography. However, existing lattice-based authentication schemes fall short of addressing the potential challenges of the leakage of the master secret key and key-escrow problem. By ingeniously addressing both issues, the paper proposes the \emph{first} quantum secure authentication scheme to eliminate the flaws while maintaining the system's overall efficiency intact. Compared to the state-of-the-art schemes, the provable security and overall performance assessment highlight the suitability of the proposed approach.    
\end{abstract}

\section{Introduction}
\begin{figure}[]
    \centering
    \includegraphics[width = \columnwidth]{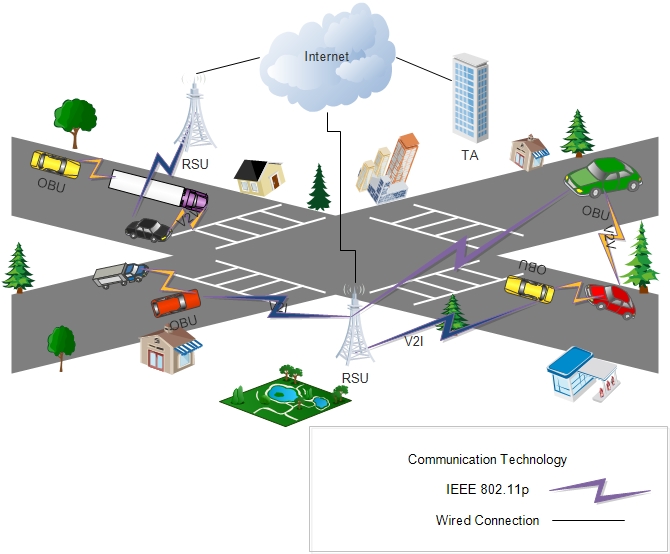}
    \caption{Basic VANET Model}
    \label{fig: Basic VANET Model}
\end{figure}

\IEEEPARstart{T}he increase in traffic has prompted the exploration of innovative solutions like ad-hoc networks and Vehicular Ad-hoc Networks (VANET). Due to the miscellaneous advancements in wireless transmission, VANETs enable nodes to engage in real-time communication with one another and with infrastructure. It allows for exchanging critical traffic-related information to positively impact traffic flow and overall transportation efficiency \cite{Lee2016an}. Classic VANET architecture has three main working components (Figure \ref{fig: Basic VANET Model}) viz: a trusted party (TA); that acts as an administrative centre and is responsible for overseeing the whole network, road-side units (RSUs); that act as intermediate nodes between vehicles and TA, and vehicles (equipped with an electronic device referred as on-board units (OBU) to receive, store and transmit the data). 

 \par In VANET, the communication among the vehicles for information sharing (like direction, location, speed, and other road statuses) is carried out by using dedicated short-range (DSRC) protocols. It is done in two ways 1) vehicle-to-vehicle communication (V2V), and 2) vehicle-to-infrastructure communication (V2I) \cite{Li2022lattice}. The V2V and V2I communications are done wirelessly, while RSUs and TA communicate through the wiring medium. Using RSUs as intermediate nodes, the information from the vehicle is sent to the TA. Then, TA analyzes and utilizes information to provide better user services. The communication of these nodes and RSUs helps avoid mishaps and provides further road services (like gas station info, parking assistance, and ambulance services) whenever required \cite{Lee2013toward}.
 
\par Although TA thoroughly monitors the network, information security is still a major problem. The reason is that the challenger can try to alter, reform, replay, or even delete the data. This wrong traffic (or route) related information or wrong traffic stat not only can cause congestion but, at some routes, it can prove fatal. In addition, a malicious node could transmit inaccurate information to the network. This can lead to possible mishaps, and these fake messages can affect the node's rights. Besides, it can cause severe innumerable problems to the overall traffic system \cite{Hubaux2004the, wenmin2012an}. Therefore, ensuring the security parameters (especially the authenticity of sources and messages) correspond to shared information is fundamental. To certify the standardization of security, the user's privacy is also essential. Thus, the simple yet choicest solution is a privacy-preserving authentication protocol to achieve authentication and privacy in a single step. However, in some cases, privacy may be conditional, i.e. in dispute, TA can reveal the user's identity. One such protocol that is capable of providing conditional privacy is the conditional privacy-preserving authentication protocol (CPPA).

\subsection{Related Work}
In 2004, Hubuax \textit{et al.} \cite{Hubaux2004the} studied various traffic-related concerns majoring in security and privacy; they also discussed the challenges correlated with smart vehicles. Later, in 2007, Raya \textit{et al.} \cite{Raya2007securing} proposed a secured architecture for vehicular ad-hoc networks and discussed the related security threats. They used certified public keys for communication, and these pre-loaded certificates protect the user's identity, although it enhanced the storage cost problem. Also, in such cases, it gets difficult for TA to extract the attacker's real identity. In the same year, Lin \textit{et al.} \cite{Lin2007GSIS} proposed an efficient pairing-based group signature scheme that provides security and privacy to users and contributes to traceability features. The utilization of the group signatures made the scheme lightweight. However, complex computing and exponentiation features reduced the overall efficiency of the scheme. A year later, Lu \textit{et al.} \cite{Lu2008ECPP} proposed an efficient conditional privacy-preserving protocol (ECPP) that provides better anonymity and privacy attributes to the users. However, a user needs a new certificate from the nearest RSU every time to communicate with another node, which ultimately leads to severe communication overhead. To lighten the drawback, Zhang \textit{et al.} \cite{zhang2008raise} introduced an innovative RSU-aided message authentication scheme (RAISE). This approach ensures user privacy through k-anonymity, inhibiting message-vehicle linkage. However, this protocol couldn't minimize the storage cost as it was again a PKI-based design.

\par Later, Zhang \textit{et al.} \cite{zhang2008an} developed a batch verification ID-based method for vehicular sensor networks. The idea aims to reduce computation through batch verification and enable the trust authority to uncover the vehicle identity associated with any pseudo-identity. However, using bilinear pairings in this approach couldn't reduce the computational cost. Later, some more improved proposals have been put forward in \cite{sun2010an, huang2011pacp, chim2011specs, shim2012cpas}, and \cite{liu2014improvements}. In \cite{liu2014proxy}, an idea of proxy vehicles has been utilized to reduce the computation burden on RSUs. In the design, some proxy vehicles share part of the computation from RSU. In \cite{Li2015acpn}, Li \textit{et al.} incorporated public-key cryptography (PKC) into pseudonym creation, enabling valid third parties to establish non-repudiation of vehicles via real ID retrieval. However, mostly the development is based on pairing computation and, therefore, is inefficient.

\par In 2015, He \textit{et al.} \cite{He2015an} proposed a pairing-free identity-based conditional privacy-preserving batch-verification protocol for vehicular ad-hoc networks. Since this protocol is pairing-free, it reduces both computational costs and communication overhead. A new approach enabling the formation of a certificateless aggregate signature (CLAS) system for securing V2I communication in VANETs has been proposed by Horng \textit{et al.} \cite{Horng2015an}. It combines certificateless public key cryptography and aggregate signature benefits, maintaining a middle ground between anonymity and traceability. It accomplishes anonymous authentication, information reliability, and unlinkability and avoids requiring synchronization for sharing the exact random string. To integrate effective tracking mechanisms, a pairing-based authentication scheme was put forward by Azees \textit{et al.} \cite{azees2017eaap} in 2017. Their method presents a robust conditional privacy tracking technique for pinpointing the identity of malicious vehicles. However, their approach overlooked the concept of non-flammability, which ensures the impossibility of creating a fake signature for a participant. Besides, in 2021, Kazemi \textit{et al.} \cite{kazemi2018on} presented an article in which they showed that the paper \cite{azees2017eaap} is not secure against modification attack, replay attack, and impersonation attack, and they also proclaim that the scheme has linkable issues. Zhang \textit{et al.} \cite{Zhang2021pa-crt} suggests a fingerprint-based verification scheme as an alternative to real identity and passwords. This scheme removes the necessity of an ideal tamper-proof device (TPD), thus preventing potential risks of compromising a vehicle's TPD and causing system malfunction. 

\par Ogundoyin \textit{et al.} \cite{Ogundoyin2020an} introduced an elliptic curve cryptography-based CPPA system for VANETs. This approach meets VANET's security and privacy needs, resolves private key compromises, and averts privilege escalation. Nonetheless, notable storage requirements arise as the network manager must retain substantial relevant data. The study \cite{haung2020secure} introduced a new privacy-protecting authentication method using 5G software-defined vehicular networks. By leveraging the benefits of this network type, the proposed method streamlines authentication to just two simple hash operations, minimizing computational work. Furthermore, it eliminates the need for the ideal TPD and avoids a vulnerable single point of failure. Security evaluation confirms that the method fulfils outlined system goals. Another certificateless aggregate signature system with conditional privacy preservation was created by Mei \textit{et al.} \cite{Mei2021efficient}. Some new and innovative developments have recently been reported in \cite{samra2022new, zhou2022efficient, xiaoxue2023ptap, bao2023bap}, and \cite{zhu2023security}.

\par All the above-mentioned schemes reflected improvements in computations and storage costs. However, none of them could resist the quantum threats \cite{shor1994algorithms}. As per the National Institute of Standards and Technology (NIST), lattice-based cryptography can be a good solution to resist quantum computing threats \cite{Li2022lattice}. Therefore, to resist quantum computing threats in VANETs authentication, Mukherjee {\it et al.} devised a lattice-based CPPA (LB-CPPA) scheme in \cite{mukherjee2019an}. The hardness of the worst-case lattice assumptions has been used to create a novel and effective LB-CPPA scheme. The devised CPPA protocol demonstrates faster computing performance and resistance to contemporary quantum computers. However, in the design, OBU holds the prime/master private key of TA. Thus, a key leakage problem is available. To enhance security by eliminating the drawback, the study \cite{Liu2019a}  employs lattice-based cryptography for an anonymous authentication scheme devoid of tamper-proof devices. This design fulfils VANET's security and privacy needs while also withstanding quantum computer attacks. However, the vehicle's secret key is computed by TA. Thus, it suffers from a key-escrow problem. As a further improvement, Dharminder \textit{et al.} \cite{Dharminder2020LCPPA} proposed a framework utilizing a short integer solution problem within a lattice. This framework accomplishes robust user identification, authorization, and service revocation in VANETs. However, the user can compute the TA's secret key from its own data. The most recent development on the line has been proposed in \cite{Li2022lattice}. The suggested scheme is secure in the ROM under the hardness of the ISIS problem. However, it suffers from master-secret key leakage from the OBU side.

\subsection{Motivation and Contribution}
As per the discussion, most of the existing LB-CPPA schemes have the drawback of master-secret key leakage, or TA computes the vehicle's secret key (Table I) (Key-Escrow). Therefore, to mitigate the issues, the CLAS paradigm can be utilized to design post-quantum CPPA schemes. In CLAS-based schemes, the secret key of the node has two parts. A part is computed by TA (called the partial secret key), and another part is picked by the vehicle itself \cite{alriyami2003certificateless}. Thus, there is no need to store the master secret of TA to OBU, and the vehicle's full secret key is also unknown to TA. Therefore, the paper aims to design the first quantum-secure certificate-less CPPA (QS-CL-CPPA) scheme. The description of the contributed scheme is as follows:

\begin{itemize}
 \item The first QS-CL-CPPA scheme has been devised in the paper. The CLAS scheme has been utilized in a lattice-based setting to support batch verification.
\item The provable security has been discussed in detail to show the robustness of the proposed QS-CL-CPPA scheme in the ROM.
 \item The informal security analysis against various attacks has been presented in detail.
 \item To prove the effectiveness of the design, a detailed computational cost and communication cost analysis has been presented.
\end{itemize}

\begin{table}[]
\caption{Comparison of the Most Related Lattice-based Work}\label{tab1}\centering
\begin{tabular}{lccc}
\hline Scheme &   SP-1 & SP-2 & SP-3\\ \hline
Li {\it et al.} \cite{Li2022lattice}&  $\times$ & $\checkmark$  & $\times$ \\ 
 Mukherjee {\it et al.} \cite{mukherjee2019an}&  $\times$ & $\checkmark$ & $\times$  \\
  Liu {\it et al.} \cite{Liu2019a} &  $\times$ & $\times$  & $\checkmark$ \\
Dharminder {\it et al.} \cite{Dharminder2020LCPPA}&  $\times$ & $\checkmark$  & $\times$ \\ 
Proposed Scheme &  $\checkmark$ & $\checkmark$ & $\checkmark$  \\ \hline
\multicolumn{4}{c}{{\bf{SP-1}}: Certificateless, {\bf{SP-2}}: No Key Escrow}\\
\multicolumn{4}{c}{{\bf{SP-3}}: No Master secret key leakage, $\checkmark$: Yes, $\times$: No }\\
 \hline
\end{tabular}
\end{table}

\par The roadmap of the proposed work is as follows:\\
Section II presents the foundation related to Lattice-theory, Network-Framework, Security Requirements, and Framework of the proposed QS-CL-CPPA scheme. The design of the proposed QS-CL-CPPA sch,eme has been put in Section III, and the security analysis of the QS-CL-CPPA scheme has been presented in Section IV. Section V presents a performance discussion, and Section VI concludes the paper. 
\section{Prelimeneries}
\subsection{Mathematical Backgroud}
To discuss Lattice-based foundation, we use the following standard definitions. For a vector $\mathbf{a}$, the Euclidean norm is defined as $\|\mathbf{a}\|=\sqrt{a_1^2+a_2^2+\cdots+a_n^2}$ where $a_1, a_2 \ldots a_n$ are the elements of vector $\mathbf{a}$. 

\begin{table}[]
\caption{Notation Table}\label{tab2} \centering

\begin{tabular}{|c|c|}
\hline Notation & Description of notations \\
\hline $k$ & Security parameter \\
$q$ &  $q $(a prime no.)  $ \approx \operatorname{poly}(k)$ \\
$m$ & $m $ (integer)$\approx O(k \log q)$ \\
$n$ & $n $ (integer)$\approx O(k \log q)$ \\
 $\mathbf{X}^{\mathbf{T}}$ & Matrix transpose $\mathbf{X}$ \\
$\|\mathbf{x}\|$ & Euclidean norm of vector $\mathbf{x}$ \\
 A & An integer matrix, $\mathbf{A} \in Z_q^{m \times n}$ \\
 d & System secret key, $\mathbf{d} \in Z_q^m$ \\
 $P_{PUB}$ & System public key, $\mathbf{P U B}=\mathbf{d}^{\mathbf{T}} \cdot \mathbf{A} \in Z_q^{1 \times n}$ \\
$H_1()$ &  $H_1: \mathbb{Z}^{1 \times n}_q \times \mathbb{Z}_q^m \times \{0, 1\}^l \longrightarrow \{0, 1\}^l$ \\
 $H_2()$ &  $H_2:\mathbb{Z}^{1 \times n}_q \times \{0,1\}^* \longrightarrow \mathbb{Z}^*_q$ \\
$H_3()$ & $H_3:\mathbb{Z}^{1 \times n}_q \times \{0,1\}^l \times \{0,1\}^* \times \{0,1\}^l \longrightarrow \mathbb{Z}^*_q$ \\
 $R_{ID_i}$ & Real ID of a node \\
 $P_{ID_i}$  & pseudo-ID of a node \\
 $ANS_i$ & An anonymous identity of $i^{\text {th }}$ vehicle \\
$m_i$ & Message of $i^{\text {th }}$ vehicle about traffic status \\
$T_i$ & Timestamp of $i^{\text {th }}$ vehicle \\
 $\mathbf{s k}_{\mathbf{i}}$ & Secret key of $i^{\text {th }}$ vehicle, $\mathbf{s k}_{\mathbf{i}} \in Z_q^m$ \\
 $\sigma_i$ & Sign of $i^{\text {th }}$ vehicle\\
$\oplus$ & XOR operation \\
$\mathbin\Vert$ & Concatenation operation \\
\hline
\end{tabular}
\end{table}

\noindent\textbf{Definition 1:} A lattice $\mathscr{L}$ is a linear combination of ${n}$, ${m}$-dimensional vectors $v_1, v_2, v_3, \ldots, v_n \in \mathbb{R}^m$. If B $=\left\{v_1, v_2, v_3, \ldots, v_n\right\}$ be a set of independent vectors, we define $\mathscr{L}({B})=\left\{x_1 v_1+x_2 v_2+\ldots+x_n v_n ; x_i \in \mathbb{Z}\right\}$ where $\mathbb{R}^m$ is a ${m}$-dimensional Euclidean space and B $=\left\{v_1, v_2, v_3, \ldots, v_n\right\}$, is a basis of $\mathscr{L}({B})$. 

\noindent\textbf{Lemma 1:}
There exists at least one basis for every lattice $\mathscr{L} \in \mathbf{R}^{{m}}$. It can be represented as a basis matrix $\boldsymbol{A}=\left[\mathbf{a}_1, \mathbf{a}_2, \ldots \mathbf{a}_{\mathbf{n}}\right] \in \mathbb{Z}^{\mathbf{m} \times {n}}$, having basis vectors as the columns.
 
\noindent\textbf{Definition 2:}
A lattice  $\mathscr{L} $ induced by a basis matrix $\boldsymbol{A} \in \mathbb{Z}^{\mathbf{m} \times {n}}$ is defined as  $\mathscr{L} (\boldsymbol{A})$ = [$\boldsymbol{A}\mathbf{s}$ : $\mathbf{s} \in \mathbb{Z}^{{n}}$ ] Where $\boldsymbol{A}\mathbf{s}$ is a multiplication of a matrix and a vector .
\par There is no unique basis $\boldsymbol{A}$ for any $\mathscr{L} (\boldsymbol{A}) $. A matrix is unimodular unimodular if the determinant is $\pm{1}$. Let $\boldsymbol{U}$ be a unimodular matrix such that $\boldsymbol{U} \in \mathbb{Z}^{\mathbf{n} \times {n}}$, then $\boldsymbol{A}$  $\boldsymbol{U}$ is a basis of  $\mathscr{L} (\boldsymbol{A}) $.
\par The minimal distance of lattice is $d_{min}(\mathscr{L})$ = min$ \left\{ \|\mathbf{a}\| : \mathbf{a} \in \mathscr{L}^* \right\}$.


\noindent\textbf{Definition 3:} {Shortest Vector Problem (SVP)} 
\par For a lattice $\mathscr{A}$ with basis $\mathscr{A} \in \mathbb{Z}^{{m} \times {n}}$, SVP seeks to find any vector ${s}$($\neq 0$ ) $ \in {X}$ such that $\|s\|={d}_{\min }({A})$, which  is computationally impracticable.

\noindent\textbf{Definition 4:} {Closest Vector Problem (CVP)} 
\par Let ${u}(\notin {B})$ be a vector and given a lattice with basis ${B} \in \mathbb{Z}^{{m} \times {n}}$, then closest vector problem seeks to find a vector ${t}$ ($\neq 0$ ) $\in {B}$ where $\|u-t\|={d}_{\min }({B})$, which  is computationally infeasible.

\noindent\textbf{Definition 5:} {Q-ary lattice} 
\par Given a lattice $\mathscr{L}$ satisfying $\mathbb{Z}_{{q}}{ }^{{n}} \subseteq \mathscr{L} \subseteq \mathbb{Z}^{{n}}$, ${q}$- an integer, where $\mathbb{Z}^{{n}}$ and $\mathbb{Z}_{{q}}^{{n}}$ are respectively the set of ${n}$ integers and set of ${n}$ integers over a finite field q. Given a matrix $A \in \mathbb{Z}_{{q}}^{m \times n}$, two categories of q-ary lattices (m-dimensional) are presented as:
$$
\Lambda_q^{\perp}=\left\{\boldsymbol{x} \in \mathbb{Z}^{{n}}: \mathbf{A} \boldsymbol{x}=\mathbf{0} (\operatorname{modq})\right\}
$$
$\Lambda_q=\left\{\boldsymbol{x} \in \mathbb{Z}^{{n}}: \boldsymbol{x}=\mathbf{A}^{{T}} \mathbf{y}\right.$ (modq), for some $\left.\mathbf{y} \in \mathbb{Z}^{{m}}\right\}$, where $\mathbb{Z}_{{q}}^{m \times n}$ is the set of matrices modulo "$q$".\\

\noindent\textbf{Definition 6:} {Small Integer Solution (SIS)}
\par Suppose $q$ be an integer, $\alpha >0$- a constant, and a matrix ${T} \in \mathbb{Z}_{{q}}^{n \times m}$, finding a vector $s$, such that ${T} s=0(\bmod {q})$ and $0<\|s\|<\alpha$ is infeasible.\\
\noindent\textbf{Definition 7:} {Inhomogeneous Small Integer Solution (ISIS)} 
\par For a matrix ${M} \in \mathbb{Z}_{{q}}^{m \times n}$, $\alpha$-an integer and a vector chosen randomly $\boldsymbol{x}\in \mathbb{Z}_{{q}}^{m}$, ISIS seeks to find a vector $\boldsymbol{y} \in \mathbb{Z}^{ n}/ \{0\} $ where  $\|\boldsymbol{y}\| < \alpha$ such that ${M} \boldsymbol{y}\  = \boldsymbol{x}(\bmod q)$ is infeasible.

\noindent\textbf{Theorem 1:}
For any polynomial bounded $\alpha>0$ and $g$ i.e. $\alpha, g=$ poly $(h)$, along with a prime modulus $q \geq \alpha \cdot \sqrt{g h}$, determining lattice problems like SIS and ISIS assumptions in the average case is always as difficult as approximating $S I V P_\gamma$ and GapSV $P_\gamma$ problems in the worst case on any n-dimensional lattice $\mathscr{L}$ within a factor $\gamma= \alpha. \tilde{O} \sqrt{h}$.\\
For a detailed discussion on these concepts, one can refer to \cite{Li2022lattice}, and \cite{mukherjee2019an}.

\begin{figure}[]
    \centering
    \includegraphics[width = \columnwidth]{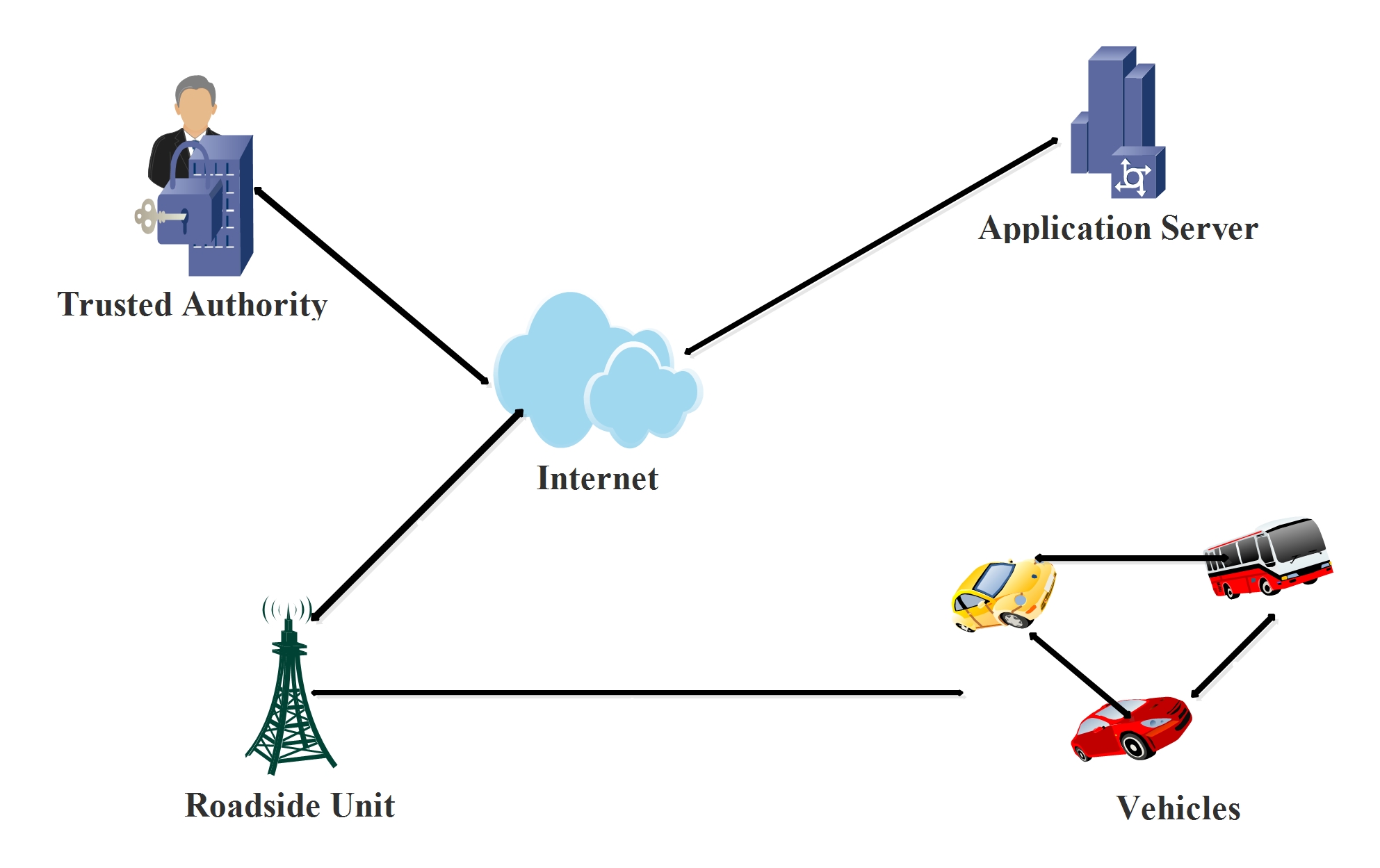}
    \caption{VANET Network Model}
    \label{fig: VANET Network Model}
\end{figure}
\subsection{Network Framework}
Generally, the VANET's network model usually adopts a dual-layer architecture and encompasses four distinct components: TA, a node/vehicle, an RSU,  and an application server (AS) as shown in Figure \ref{fig: VANET Network Model}. A detailed breakdown of these entities follows:

\begin{itemize}
    \item TA:
The trusted third authority serves the role of ensuring the integrity and security of communications within the VANET. It authenticates vehicles, verifies their identities, and grants various actions and services permissions. The TA also manages cryptographic keys, enabling secure message exchanges and preventing unauthorized access.
\item AS: The application server acts as a central point for service delivery, data management, communication, security, and overall coordination of applications to enhance the functionality and effectiveness of the vehicular network.
\item RSU: These roadside units are planted along the roadsides, and they work as a transitional node connecting the TA and the nodes. They perform many functions, including communication, data collection, traffic control, emergency services, and infrastructure support, all of which collectively enhance road accident prevention, traffic regulation, and the overall efficiency of vehicular networks.
\item Vehicle: Each node within the VANET necessitates an OBU. The OBU functions as an impenetrable device, ensuring information remains confidential. Moreover, it enables wireless transmission between a node and an RSU or other nodes, adhering to the DSRC protocol.
\end{itemize}

\subsection{Security Requirements}
Security and anonymity in VANETs are paramount due to the delicate nature of the information exchanged and the potential impact of compromised communications. Various aspects, including unlinkability, information authentication, anonymity preservation, conditional traceability, and resistance to attacks, are essential components that ensure the robustness and reliability of the VANETs.

\begin{itemize}
    \item Information Authentication:
It ensures that messages are not tampered with or fabricated during transmission. This is crucial for maintaining the integrity of critical information such as traffic updates, emergency alerts, and vehicle coordination. TA and RSUs play pivotal roles in verifying the authenticity of messages, using cryptographic techniques to validate the source and content of messages.
\item Anonymity Preservation:
Preserving the privacy of individual vehicle identities is a significant concern in VANETs. RSUs and TAs collaborate to employ techniques like pseudonyms, which replace actual vehicle identifiers with temporary ones. This shields the real identities of vehicles from being exposed to potential eavesdroppers while still allowing legitimate authorities to authenticate vehicles.
\item Conditional Traceability:
Tracing and attributing actions back to specific vehicles is important for accountability and security. TAs and RSUs manage cryptographic keys and pseudonyms, enabling conditional traceability when required for incident investigation or law enforcement purposes.
\item Un-linkability:
Un-linkability ensures that different messages from similar nodes cannot be linked together, enhancing privacy. RSUs and TAs implement methods to prevent unauthorized entities from connecting pseudonyms to real identities, safeguarding the unlinkability of messages.
\item Resistance to Attacks:
VANETs are inclined to various quantum attacks, including man-in-the-middle attacks, modification attacks, impersonation attacks, and replay attacks.
\end{itemize}

 \subsection{Framework of the Proposed QS-CL-CPPA Scheme}
 The proposed QS-CL-CPPA scheme is comprised of seven phases: System Initialization, $P_{ID}$-Gen, psk-Gen, sk-Gen, $CL_{sig}$-Gen,  $CL_{sig}$-Ver, and Batch-$CL_{sig}$-Ver. The detailed execution is as follows:
\begin{itemize}

\item System Initialization: This phase is run by TA with a security parameter as input. It is also known as the system setup phase as all system parameters $\left\{m, n, q, \mathbf{A}, P_{PUB}, H_1, H_2, H_3\right\}$ are generated and kept in the OBU's and the existing RSU's by TA. 

\item $P_{ID}$-Gen: In this phase, every vehicle ($V_i$) interacts with the TA to generate the pseudo-identity ($P_{ID_i}$). After certain computations, TR sends $P_{ID_i}$ and time stamp $T_i$ to the vehicle. Then, TA and vehicle store the data. 
       
\item psk-Gen: In the phase, TA generates the partial-secret key ($psk_{ID_i}$)  of $P_{ID_i}$ and shares it secretly with the vehicle ($P_{ID_i}$).

\item sk-Gen: After obtaining the partial-secret key from TA, in this phase, vehicle ($P_{ID_i}$) computes its full secret key and outputs its public key $pk_{ID_i}$.

\item $CL_{sig}$-Gen: For the signature generation phase, the vehicle ($P_{ID_i}$) outputs a signature $\sigma_i$ while taking message $m_i$ and secret-key $sk_{ID_i}$ as input.
    
\item $CL_{sig}$-Ver: While the message is received by the RSU/OBU, the authentication for the signature is carried out. Here, on inputting the signature $\sigma_i$ of the vehicle ($P_{ID_i}$), the output turns out true only if $\sigma_i$ is valid, else it will return false

\item Batch-$CL_{sig}$-Ver: In this phase, "$n$" a number of message-signature pairs is taken as input by the RSU's/OBU's and they return true or false accordingly if the signatures are valid or not.
\end{itemize}

\section{Proposed QS-CL-CPPA Scheme}
In this section, the proposed QS-CL-CPPA scheme has been presented in detail. All the notations/symbols used in the suggested scheme are noted in Table \ref{tab2}
\begin{itemize}
\item System initialization:
The trusted authority TA brings out the system parameters, which are then loaded into the secure devices of individual mobile nodes beforehand. Subsequently, these parameters are transmitted to all Roadside Units (RSUs). Choosing \textit{k} as a security parameter, TA takes $ 1^\textit{k}$ as input and outputs the subsequent parameters:\\
(a) The TA takes two integers $m, n$, and a prime number $q$.\\
(b) The TA chooses a matrix $A \in Z_q^{m \times n}$.\\
(c) A vector $d \in Z_q^m$ is chosen randomly as the system private key by the TA and computes the system public key as $P_{Pub}=d^T A \in Z_q^{1 \times n}$.\\
(d) TA chooses the one-way hash functions as; $H_1: \mathbb{Z}^{1 \times n}_q \times \mathbb{Z}_q^m \times \{0, 1\}^l \longrightarrow \{0, 1\}^l$, $H_2:\mathbb{Z}^{1 \times n}_q \times \{0,1\}^* \longrightarrow \mathbb{Z}^*_q$, and $H_3:\mathbb{Z}^{1 \times n}_q \times \{0,1\}^l \times \{0,1\}^* \times \{0,1\}^l \longrightarrow \mathbb{Z}^*_q$.\\
(e) The final system parameters $\triangle=\left\{q, m, n,  P_{Pub}, A, H_1, H_2, H_3\right\}$ are distributed  among RSUs and vehicles.

\item $P_{ID}$- Gen: 
Vehicle and TA interact to generate $P_{ID_i}$ (pseudo-identity) of the vehicle's real- identity $R_{ID_i}$ using secure-channel.\\
(a) Vehicle takes $a_i \in_R \mathbb{Z}_q^m$  vector and computes $ pk^1_{ID_i} = a_i^T A $ and shares $(pk^1_{ID_i}, R_{ID_i})$ to TA.\\
(b) TA Computes $P_{ID_i} = R_{ID_i}$ $\oplus$  $H_1(pk^1_{ID_i}, d, T_i)$ and sends $P_{ID_i}$ and $T_i$ (time-stamp) to vehicle.\\
(C) TA and vehicle both store $ID_i$ = ($pk^1_{ID_i}, P_{ID_i}, R_{ID_i}, T_i)$.
        
\item psk- Gen:
TA executes the following steps to generate the partial-private/secret key ($psk_{ID_i}$) of the vehicle (with $ID_i$).\\
(a) Picks $x_i$ $\in_R$ $Z_q^m$ vector and computes $X_i = x_i^T A$.\\
(b) Computes $psk_{ID_i}$=$x_i+\gamma_i d$, where $\gamma_i = H_2(pk^1_{ID_i} + X_i, T_i)$ and TR shares $(psk_{ID_i}, X_i)$ to vehicle ($ID_i$) secretly.

\item sk-Gen:
In this phase, the vehicle does the following steps:\\
(a) Computes $sk_{ID_i}$ = $psk_{ID_i} + a_i$ as the secret-key.\\
(b) Outputs $pk_{ID_i}$ = $(pk^1_{ID_i}+ X_i)$ as the public-key.\\
Public key is broadcasted to the nearby atmosphere.
    
\item $CL_{sig}$-Gen:
To broadcast a safety-related message $m_i$, vehicle $(ID_i)$ generates a signature as follows:\\
(a) Selects $r_i \in_R Z_q^m$ vector and computes $R_i$ = $r_i^T A$\\
(b) Computes $\sigma_i$=$sk_{ID_i}+ \delta_i r_i$, where $\delta_i = H_3(R_i, P_{ID_i}, m_i, T_i)$.\\
Then, vehicle $(ID_i)$ broadcasts ($m_i, \sigma_i, R_i$) to nearby OBUs and RSUs.

\item $CL_{sig}$-Ver:
To verify the authenticity of received $(m_i, \sigma_i)$, RSU/OBU does the following steps:\\
(a) Computes $\gamma_i$ = $H_2(pk_{ID_i}, T_i)$ and $\delta_i= H_3(R_i, P_{ID_i}, m_i, T_i)$.\\
(b) Checks $\sigma_i^T A = pk_{ID_i} + \gamma_i P_{Pub} + \delta_i R_i$. If valid, outputs are accepted. Otherwise, outputs are rejected.
    
\item Batch-$CL_{sig}$-Ver:
To verify the authenticity of the "$n$" message-signature pairs, RSU/OBU performs the following:\\
(a) Computes $\gamma_i = H_2(pk_{ID_i}, T_i)$ and $\delta_i = H_3(R_i, P_{ID_i}, m_i, T_i)$, $1\leq i \leq n$ and selects $b_1, b_2,..., b_n$ $\in_R$ $\mathbb{N}$ (where $b_i's$ are samll).\\
(b) Checks $(\sum\limits_{i=1}^{n} b_i \sigma_i)^T A$ = $\sum\limits_{i=1}^{n} b_i pk_{ID_i} + (\sum\limits_{i=1}^{n} \gamma_i b_i) P_{Pub} + \sum\limits_{i=1}^{n} b_i \delta_1 R_i$.

\end{itemize}

\section{Analysis of QS-CL-CPPA Scheme}
\subsection{Correctness}
Correctness of $CL_{sig}-Ver$ and Batch-$CL_{sig}-Ver$ can be seen as follows:\\
(a) $$\sigma_i^T A  = (sk_{ID_i} + \delta_i r_i)^TA$$
$$ = (psk_{ID_i} + a_i)^TA + \delta_i r_i^T A$$
$$ = (x_i + a_i + \gamma_i d)^TA + \delta_i R_i $$
$$ = x_i^TA + a_i^TA + \gamma_i d^TA + \delta_i R_i$$
$$  = X_i + pk^1_{ID_i}  + \gamma_i P_{Pub} + \delta_i R_i$$
$$  =  pk_{ID_i}  + \gamma_i P_{Pub} + \delta_i R_i$$\\

(b) $$(\sum\limits_{i=1}^{n} b_i \sigma_i)^T A = \sum\limits_{i=1}^{n} b_i (sk_{ID_i} + \delta_i r_i)^TA$$
$$ =  \sum\limits_{i=1}^{n} b_i sk_{ID_i}^T A + \sum\limits_{i=1}^{n} b_i \delta_i r_i^TA$$
$$ = \sum\limits_{i=1}^{n} b_i (psk_{ID_i} + a_i)^T A + \sum\limits_{i=1}^{n} b_i \delta_i r_i^TA$$
$$ = \sum\limits_{i=1}^{n} b_i (x_i + \gamma_i d + a_i)^T A + \sum\limits_{i=1}^{n} b_i \delta_i r_i^TA$$
$$ = \sum\limits_{i=1}^{n} b_i (x_i + a_i)^T A + \sum\limits_{i=1}^{n} b_i \gamma_id^TA + \sum\limits_{i=1}^{n} b_i \delta_i r_i^TA$$
$$ = \sum\limits_{i=1}^{n} b_i pk_{ID_i} + (\sum\limits_{i=1}^{n} b_i \gamma_i) P_{PUB} + \sum\limits_{i=1}^{n} b_i \delta_i R_i$$

\subsection{Provable security Analysis}
\begin{itemize}
\item[A]\textbf{Theorem-2}: if the SIS/ISIS problem in lattice theory is interactable, the proposed QS-CL-CPPA Scheme is secure against adversaries $\mathcal{A}_1$and $\mathcal{A}_2$ in the ROM.\\
\noindent\textbf{Proof:} The proof of the theorem can be attained from Lemma-1 and Lemma-2 as;
    
\item[B] \textbf{Lemma-1}: If the SIS/ISIS problem in lattice theory is interactable, the proposed QS-CL-CPPA Scheme is secure against $\mathcal{A}_1$ adversary in ROM.\\
\textbf{Proof}: Suppose $\mathcal{A}_1$ can forge the underlying signature scheme. Now, the following interactive game involves $\mathcal{A}_1$ and challenger $\mathcal{C}$.  
    
 \item \textbf{Setup()}: $\mathcal{C}$ generates parameters $\triangle$ and shares to $\mathcal{A}_1$. $\mathcal{A}_1$ cannot get the secret key of TR. $\mathcal{C}$ maintains lists $L_{H_2}$, $L_{H_3}$, $L_{pk}$ , $L_{psk}$ and $L_{sk}$ initially empty.
    
 \item \textbf{Queries $(H_2)$}: $\mathcal{A}_1$ puts $P_{{ID}_i}$ to $H_2(.)$ oracle. $\mathcal{C}$ scans $L_{H_2}$, if the entry corresponding to $P_{{ID}_i}$ is available, picks the entry, and sends it to  $\mathcal{A}_1$. Otherwise, $\mathcal{C}$ picks $\gamma_i \in_R \mathbb{Z}_q^*$ and sets
 $\gamma_i$=$H_2$ $(pk_{ID_i}, T_i)$.  $\mathcal{C}$ update $L_{H_2} \cup (P_{{ID}_i}, \gamma_i)$ and sends $\gamma_i$ to $\mathcal{A}_1$.

 \item \textbf{Queries $(H_3)$}: $\mathcal{A}_1$ puts $(P_{{ID}_i}, m_i , R_i, T_i)$ to  $H_3(.)$ oracle. $\mathcal{C}$ scans $L_{H_3}$ and if an entry is found, $\mathcal{C}$ sends the entry to $\mathcal{A}_1$. Otherwise, $\mathcal{C}$ picks $\delta_i \in_R Z_q^*$ and sets $\delta_i = H_3(R_i, P_{ID_i}, m_i, T_i)$. $\mathcal{C}$ updates $L_{H_3}\cup (P_{{ID}_i}, R_i, m_i, T_i, \delta_i)$ and sends $\delta_i$ to $\mathcal{A}_1$.

 \item \textbf{Reveal psk-Queries}: $\mathcal{A}_1$ requests $P_{{ID}_i}$ to psk-Gen(.) oracle. $\mathcal{C}$ first searches $L_{psk}$ to get an entry associated with $ P_{{ID}_i}$. If the entry corresponding to $P_{{ID}_i}$ exists, $\mathcal{C}$ shares the entry to $\mathcal{A}_1$. Otherwise, $\mathcal{C}$ does the following:\\
- $\mathcal{C}$ picks $psk_{{ID}_i} \in_R Z_q^m$ and computes $x_i = psk_{{ID}_i} - \gamma_i d$ and $X_i = x_i^T A$, where $\gamma_i$ taken from $L_{H_2}$. Sets the output $(psk_{ID_i}, X_i)$ and update $L_{psk}\cup\{P_{{ID}_i},psk_{{ID}_i}, x_i, X_i\}$ and sends $(psk_{{ID}_i}, X_i)$ to  $\mathcal{A}_1$.

 \item \textbf{Reveal sk-Queries}: $\mathcal{A}_1$ requests $P_{{ID}_i}$ to sk-Gen(.) oracle. $\mathcal{C}$ first checks $L_{sk}$ to get an entry corresponding to $P_{{ID}_i}$. If the entry exists, $\mathcal{C}$ picks and sends the entry to $\mathcal{A}_1$. Otherwise, do the following:\\
(a) $\mathcal{C}$ picks $(P_{{ID}_i},psk_{{ID}_i}, x_i, X_i)$ from $L_{psk}$ and picks $a_i \in_R Z_q^m$ and computes $pk^1_{{ID}_i} = a_i^T A$.\\
(b) At last, $\mathcal{C}$ update $L_{sk}\cup \{P_{{ID}_i},psk_{{ID}_i}, x_i, X_i, pk^1_{{ID}_i}, a_i\}$ and outputs $sk_{{ID}_i}$ equal to $psk_{{ID}_i} + a_i$ as secret key.

\item \textbf{Reveal $pk_{{ID}_i}$-Queries}:  $\mathcal{A}_1$ puts  $P_{{ID}_i}$ to get associated with $pk_{{ID}_i}$. to respond, $\mathcal{C}$ acts as follows:\\
(a) $\mathcal{C}$ scans $L_{pk}$ and if an entry exits, picks the entry ($pk_{{ID}_i}$) and sends it to $\mathcal{A}_1$.\\
(b) Otherwise, $\mathcal{C}$ runs psk-Gen(.) and sk-Gen(.) oracle by $P_{{ID}_i}$ as input and get  $(psk_{{ID}_i}, x_i, X_i)$  then sets $ pk_{{ID}_i}$=$pk^1_{{ID}_i} + X_i $ as public key and update 
$L_{pk}$ by adding \{$P_{ID_i}, pk_{{ID}_i}$\}.

\item \textbf{Replace $pk_{{ID}_i}$-Queries}: $\mathcal{A}_1$ puts \{ $P_{ID_i}, pk'_{{ID}_i}$\} to replace the $ pk_{{ID}_i}$, while associate secret key is not known to $\mathcal{A}_1$. $\mathcal{C}$ replace the old value in $L_{pk}$.

\item \textbf{$Cl_{{sig}}$-Gen Queries}: $\mathcal{A}_1$ puts \{$m_i, P_{ID_i}$\} to $Cl_{{sig}}$-Gen(.) oracle to obtain a sign on $m_i$. Challeneger $\mathcal{C}$ picks \{$P_{ID_i}$, $\gamma_i$\} $\in$ $L_{H_2}$ and
$\{P_{ID_i}, m_i, R_i, T_i, \delta_i\}$ $\in$ $L_{H_3}$. Then $\mathcal{C}$ picks $\sigma_i$ $\in$ $Z_q^m$ and sets $pk_{{ID}_i}$ = $\sigma_i^TA$-$\gamma_i P_{Pub}-\delta_i R_i$ and outputs $\{ m_i, \sigma_i\}$ as a message signature pair to $\mathcal{A}_1$. Everyone can verify that $\sigma_i^T A$ = $pk_{{ID}_i}$+$\gamma_i P_{Pub}+\delta_i R_i$. Thus $\{ m_i, \sigma_i\}$ simulated by $\mathcal{C}$ is indistinguishable from the real one.
\item \textbf{Forgery}: Eventually, forger $\mathcal{A}_1$ outputs a forged $\{ m_i^*, \sigma_i^*\}$ for a targeted $p^*_{{ID}_i}$ such that $\sigma_i^{*T}A$= $pk^*_{{ID}_i}$+$\gamma^*_i P_{Pub}+\delta^*_i R_i$ holds and $P^*_{{ID}_i}$ has never been requested to psk-Gen(.) oracle and $m_i^*$ has never been put to $Cl_{{sig}}$-Gen oracle.\\
 By Forking Lemma \cite{Bellare2019the}, $\mathcal{A}_1$ can output another signature $\{ m_i^*, \sigma_i^{**}\}$ for the targeted $P^*_{{ID}_i}$ with different $\sigma_i^{**}$ which satisfies\\
$\sigma_i^{**T}A =  pk^*_{{ID}_i} + \gamma^{**}_i P_{Pub}+\delta^*_i R_i$\\
$\implies (\sigma_i^* - \sigma_i^{**})^TA = (\gamma_i^* - \gamma_i^{**})d^T A$\\
$\implies (\sigma_i^* - \sigma_i^{**}) = (\gamma_i^* - \gamma_i^{**})$ $d$ mod $q$\\
Thus, $(\gamma_i^* - \gamma_i^{**})^{-1}(\sigma_i^* - \sigma_i^{**})$ mod $q$ is a solution to ISIS problem with instance $P_{Pub}$ = $d^TA$, which is a contradiction to the assumption.
\\
Thus, the proposed $Cl_{{sig}}$-Gen is unforgeable under the intractability of the SIS/ISIS problem.
\item [C] \textbf{Lemma 2}: If the SIS/ISIS problem in lattice theory is interactable, the proposed QS-LC-CPPA is secure against $\mathcal{A}_2$ adversary in ROM.\\
\textbf{Proof}: $\mathcal{C}$ generates parameters $\triangle$ shares to 
$\mathcal{A}_2$. Besides, $\mathcal{C}$ gives the master secret key of TR to $\mathcal{A}_2$. However, $\mathcal{A}_2$ cannot get a user's secret key. $\mathcal{C}$ maintains the lists  $L_{H_2}$, $L_{H_3}$, $L_{pk}$ , $L_{psk}$ and $L_{sk}$ initially empty. Similar to lemma 1, $\mathcal{A}_2$ requests $H_2(.)$ and $H_3(.)$ oracles and get the outputs from $\mathcal{C}$. However, the other queries are put by $\mathcal{A}_2$ in the following manner:
\item \textbf{Reveal $psk_{{ID}_i }$- Queries}: $\mathcal{A}_2$ puts $P_{{ID}_i}$ to psk-Gen(.) oracle and $\mathcal{C}$ responds as follows:\\
(a) $\mathcal{C}$ scans $L_{psk}$ to get entry corresponding to $P_{{ID}_i}$, if tuple $(P_{{ID}_i}$,$psk_{{ID}_i}$, $x_i$, $X_i$) exists, $\mathcal{C}$ shares ($psk_{{ID}_i}$, $X_i$) to $\mathcal{A}_2$.\\
(b) Otherwise, if $P_{{ID}_i}$ $\neq$ $P^*_{{ID}_i}$, $\mathcal{C}$ picks $P_{{ID}_i}\in_R \mathbb{Z}_q^m$ and computes $x_i$=$psk_{{ID}_i}$-$\gamma_id$ and $X_i$ = $x_i^TA$, here $\gamma_i \in L_{H_2}$. $\mathcal{C}$ shares $(psk_{{ID}_i}$, $X_i$) to $\mathcal{A}_2$ and adds ($P_{{ID}_i}$ ,$psk_{{ID}_i}$,  $x_i$, $X_i$) to $L_{psk}$.\\
(c) If $P_{{ID}_i}$=$P_{{ID}^*_i}$, $\mathcal{C}$ responds $\perp$ and abort.
\item \textbf{Reveal $sk_{{ID}_i }$- Queries}:  $\mathcal{A}_2$ puts $P_{{ID}_i}$ to sk-Gen(.) oracle and obtains response as follows:\\
(a) $\mathcal{C}$ searches $L_{sk}$ to obtain a tuple ($P_{{ID}_i}$ ,$psk_{{ID}_i}$,  $x_i$, $X_i$, $pk^1_{{ID}_i}$, $a_1$). If the tuple exists, response to $\mathcal{A}_2$ is $sk_{{ID}_i}$ = $psk_{{ID}_i}$+$a_1$.\\
(b) Otherwise, If $P_{{ID}_i} \neq P_{{ID}^*_i}$, $\mathcal{C}$ obtains ($P_{{ID}_i}$, $psk_{{ID}_i}$, $x_i$, $X_i$)$\in L_{psk}$ and selects $a_i$ $\in_R$ $Z^m_q$. Then, computes $sk_{{ID}_i}$ = $psk_{{ID}_i}$+$a_1$ and $pk^1_{{ID}_i}$=$a^T_i A$. $\mathcal{C}$ adds ($P_{{ID}_i}$, $psk_{{ID}_i}$, $x_i$, $X_i$, $pk^1_{{ID}_i}$, $a_i$) to $L_{sk}$ and shares $sk_{{ID}_i}$ to $A_2$.\\
(c) If $P_{{ID}_i}$=$P_{{ID}^*_i}$ , C responds $\perp$ and abort.
\item \textbf{Reveal $pk_{{ID}_i}$-Queries}: $\mathcal{A}_2$ puts $P_{{ID}_i}$ as input and $\mathcal{C}$ searches the entry ($P_{{ID}_i}$, $pk_{{ID}_i}$)$\in L_{pk}$, $\mathcal{C}$ picks it and shares wit $\mathcal{A}_2$. Otherwise, does the following:\\
(a) If $P_{{ID}_i} \neq P_{{ID}^*_i}$, $\mathcal{C}$ runs $psk_{{ID}_i}$-Gen(.) and $sk_{{ID}_i}$-Gen(.) oracles to get ($psk_{{ID}_i}$, $pk^1_{{ID}_i}$, $X_i$) and then sets  $pk_{{ID}_i}$=$pk^1_{{ID}_i}$+$X_i$ as the public key and adds ($P_{{ID}_i}$, $pk_{{ID}_i}$) to $L_{pk}$ and share to $\mathcal{A}_2$.\\
(b) If $P_{{ID}_i}$=$P_{{ID}^*_i}$, $\mathcal{C}$ picks $x_i, y_i \in Z^m_q$ and computes $X_i=x_i^TA$, $pk_{ID_i}=y^T_iA$ and sets $pk^1_{ID_i}=pk_{ID_i}-X_i$. Then, $\mathcal{C}$ chooses $\gamma_i \in_R \mathbb{Z}_q^*$ and sets $\gamma_i$=$H_2(pk_{{ID}_i}$, $T_i$). Finally, $\mathcal{C}$ adds ($pk_{{ID}_i}$, $pk^1_{{ID}_i}$, $\perp$, $\gamma_i$, $X_i$) to $L_{pk}$ and shares $pk_{{ID}_i}$ to  $\mathcal{A}_2$.
\item \textbf{$Cl_{{sig}}$-Gen Queries}: $\mathcal{A}_2$ puts \{ $m_i, P_{ID_i}$\} to get signature. $\mathcal{C}$ responds as follows:\\
(a) If $P_{{ID}_i} \neq P_{{ID}^*_i}$, Challeneger $\mathcal{C}$ picks \{ $P_{ID_i}$, $\gamma_i$\} $\in$ $L_{H_2}$ and
$\{P_{ID_i}, m_i, R_i, T_i, \delta_i\}$ $\in$ $L_{H_3}$. Then $\mathcal{C}$ picks $\sigma_i$ $\in$ $Z_q^m$ and sets $pk_{{ID}_i}$ = $\sigma_i^TA$-$\gamma_i P_{Pub}-\delta_i R_i$ and outputs $\{ m_i, \sigma_i\}$ as a message signature pair to $\mathcal{A}_2$. Everyone can verify that $\sigma_i^T A$ = $pk_{{ID}_i}$+$\gamma_i P_{Pub}+\delta_i R_i$. Thus $\{ m_i, \sigma_i\}$ simulated by $\mathcal{C}$ is indistinguishable from the real one.\\
(b) If $P_{{ID}_i}$ = $P_{{ID}^*_i}$, $\mathcal{C}$ responds $\perp$ and abort.
\item \textbf{Forgery}:
Eventually, $\mathcal{A}_2$ outputs a forgery \{$m_i^*, \sigma^*_i$\} for a target $P_{{ID}^*_i}$ such that $\sigma^{*T}_iA$ = $P_{{ID}^*_i}$ + $\sigma^*_i$ $P_{Pub}$ + $\delta^*_i$ $R_i$ holds and  $P_{{ID}^*_i}\notin sk$-Gen(.) and $\{P_{{ID}^*_i}, m_i\}$ $\notin$ $CL_{sig}$-Gen(.).\\
By Forking lemma \cite{Bellare2019the}, $\mathcal{A}_2$ can output another signature $\{m_i^*, \sigma_i^{**}\}$ for the targeted $P_{{ID}^*_i}$ with different $\sigma_i^{**}$, which satisfies $$ \sigma_i^{**T}A = Pk_{{ID}^*_i} + \gamma_i^* P_{Pub} + \sigma_i^{**} R_i$$
$$\implies (\sigma_i^{*}- \sigma_i^{**})^T A = (\sigma_i^{*}- \sigma_i^{**}) R_i = (\delta_i^* - \delta_i^{**}) r_i^T A$$
$\implies r_i = (\delta_i^* - \delta_i^{**})^{-1} (\sigma_i^{*}- \sigma_i^{**})$ mod $q$, 
which is a solution to ISIS problem with instance $(R_i = r_i^T A, A)$, i.e., a contradiction has been obtained.
\par Thus, the suggested scheme $CL_{sign}$ is unforgeable under the hardness assumption of the SIS/ISIS problem.
\end{itemize}

\subsection{Security Analysis}
This segment introduces a thorough discussion of informal security analysis.
\begin{enumerate}
    \item \textbf{Authentication/Integrity}: From the proof of Lemma 1 and Lemma 2, it has been justified that the proposed QS-CL-CPPA scheme is strongly unforgeable against stronger adversaries $\mathcal{A}_1$ and $\mathcal{A}_2$. Thus, only legitimate vehicles can create a signature on safety-related information. Therefore, the receiver is convinced about the authenticity of the information and source. As the signature verification can be checked by $\sigma^T A = pk_{ID} + \gamma P_{Pub} + \delta R$. Thus, modification to the message will result in the rejection of signature verification. Therefore, the integrity of the message has been achieved.

     \item \textbf{Anonymtiy}: During $P_{ID}-Gen(.)$, the real ID $(R_{ID})$ of the node is encrypted as $P_{ID}$ = $R_{ID} \oplus H_1(pk^1_{ID}, d, T_i)$. To recover real identity $(R_{ID})$, the value $H_1(pk^1_{ID}, d, T)$ is required. Here, $d$ is the master secret key of TA. Thus, only TA can reclaim any node's real ID $(R_{ID})$. Therefore, the proposed QS-CL-CPPA scheme is anonymous.
     
      \item \textbf{Conditional Traceability}: As per the discussion, real ID $(R_{ID})$ of a node is anonymised by computing $P_{ID}$=$R_{ID} \oplus H_1(pk^1_{{ID}_i}, d, T)$. Thus, in case of dispute or accident, TA can find $(R_{ID})$=$P_{ID} \oplus H_1(pk^1_{ID}, d, T)$ as $d$ is the master private key of TA. Therefore, the proposed QS-CL-CPPA scheme provides conditional traceability.
      
       \item \textbf{Resistance Against Various Attacks}:
       \item[*] \textbf{Modification Attack}: According to the discussion, to verify the authenticity of (message, signature) pair, the receiver has to check $\sigma^T A = pk_{ID} + \gamma P_{Pub} + \delta R$. Thus, modification to the message/signature will result in verification rejection. Therefore, the proposed QS-CL-CPPA scheme is robust against this attack.

       \item[*] \textbf{Impersonation attack}:  To impersonate the signing process, the attacker has to create a forgery on a targeted message/identity. However, proof of Lemma 1 and Lemma 2 shows that the QS-CL-CFPA scheme is unforgeable under SIS/ISIS problem. Thus, impersonation tack is unsuccessful.

\item[*] \textbf{Replay attack}: Based on the design of the construction, time stamps have been utilized during $P_{ID}$ and signature generation. Thus, the freshness of the timestamp can be detected to resist this attack.   
\end{enumerate}

\section{Performance Analysis}
The rapid change in the VANET topology necessitates enhanced performance concerning computational cost and communication complexity. Thus, this section details the performance analysis of the proposed QS-CL-CPPA scheme by comparison with the most related state-of-the-art schemes \cite{Li2022lattice, mukherjee2019an, Liu2019a}, and \cite{Dharminder2020LCPPA}. To achieve the security level of 123 bits corresponding to SIS/ISIS problem, the basic setup parameters have been taken from \cite{Li2022lattice}. Thus, module $q$ is set at 101, and row and column numbers of the master public key are 100 and 666, respectively, where $A \in Z ^{m \times n}$ is the master public key. 

\subsection{Compuational cost Analysis}
Since the work by Li {\it et al.}'s \cite{Li2022lattice} has been followed, the execution time of various basic cryptographic operations is provided in Table \ref{tab3}. Each operation is defined as:\\

\noindent $E_{\text {add-$v_k$ }}$: Execution time of vector addition of dimension k in $\mathbb{Z}_q$. \\
$E_{\text {add-$v_n$ }}$: Execution time of vector addition of dimension n in $\mathbb{Z}_q$.  \\
$E_{\text {add-$v_m$ }}$: Execution time of vector addition of dimension m in $\mathbb{Z}_q$.  \\
$E_{\text {mul-$v_q$cx}}$: Execution time of d * x mod q where vector x $\in \mathbb{Z}_q^m$  and d $\in \mathbb{Z}_q$.  \\
$E_{\text {mul-$v_q$dy}}$: Execution time of c * y mod q where vector y $\in \mathbb{Z}_q^n$  and c $\in \mathbb{Z}_q$. \\
$E_{\text {mul-$v_q$A}}$: Execution time of mul of a vector and matrix(nxm) on mod q. \\
$E_{\text {mul-$v_q$xA}}$: Run time of $A.x$ mod $q$, where $x \in \mathbb{Z}_q^k$ and matrix A $\in \mathbb{Z}_q^{m \times k} $  \\
$E_{\text {mul-$v_q$yA}}$: Execution time of $Ay$ mod $q$, where $y \in \mathbb{Z}_q^k$ and matrix $A \in \mathbb{Z}_q^{n \times k} $  \\
$E_{\text {mul-mat }}$: Execution time of  matrix multiplication on mod $q$. \\
$E_{\text {h}}$: Denotes the hashing function's run time.\\
$E_{\text {xor}}$: Displays the run time for performing the XOR operation of two vectors.\\
$E_{\text {sp}}$: Execution time of sampling the random vector. \\
$E_{\text {sol-eq }}$: Execution time of the equation $AB=0$ mod $q$, where $B$ is the solution of the equation set.\\

\begin{table}[]
\caption{Cryptographic Operations with Execution time }\label{tab3}\centering
\begin{tabular}{|c|c|}
\hline Cryptographic Operation & Execution Time \break (in milliseconds) \\
\hline$E_{\text {add-$v_k$ }}$ & 0.47 \\
\hline$E_{\text {add-$v_n$ }}$  & 0.059 \\
\hline$E_{\text {add-$v_m$ }}$  & 0.385 \\
\hline$E_{\text {mul-$v_q$cx}}$ & 0.697 \\
\hline$E_{\text {mul-$v_q$dy}}$ & 0.074 \\
\hline$E_{\text {mul-$v_q$A}}$ & 48 \\
\hline$E_{\text {mul-$v_q$xA}}$ & 6 \\
\hline$E_{\text {mul-$v_q$yA}}$ & 28 \\
\hline$E_{\text {mul-mat }}$ & 419 \\
\hline$E_{\text {h}}$ & 0.006 \\
\hline$E_{\text {xor}}$ & 0.0002 \\
\hline$E_{\text {sp}}$ & 15 \\
\hline$E_{\text {sol-eq }}$ & 28000 \\
\hline
\end{tabular} \end{table}

\begin{figure}[h]
\centering
\begin{tabular}{c}
\includegraphics[width= \columnwidth]{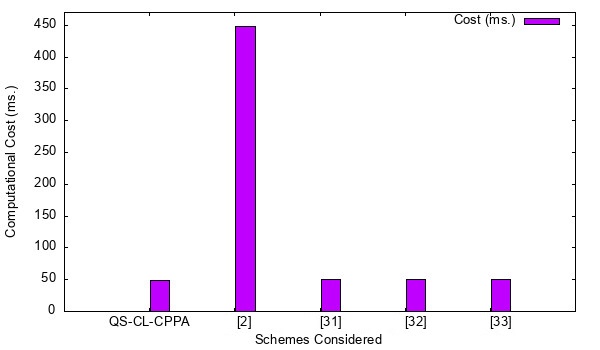}
\end{tabular}
\caption{Computational cost Sign (OBU)}
\label{fig4}
\end{figure}

\begin{figure}[h]
\centering
\begin{tabular}{c}
\includegraphics[width= \columnwidth]{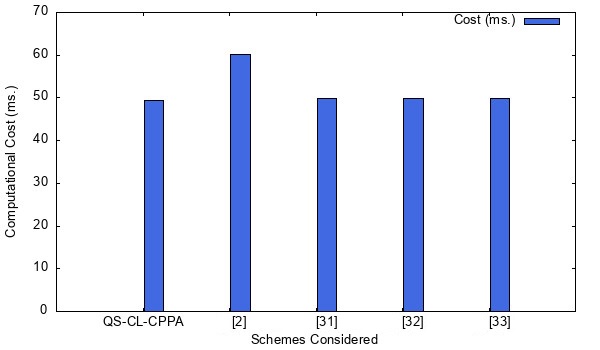}
\end{tabular}
\caption{Computational cost Single Sign-Verify}
\label{fig5}
\end{figure}

\begin{figure}[h]
\centering
\begin{tabular}{c}
\includegraphics[width= \columnwidth]{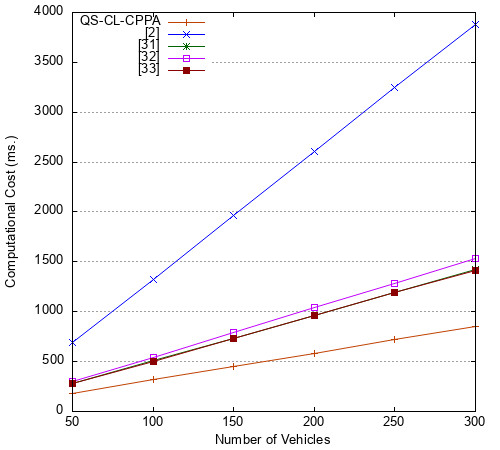}
\end{tabular}
\caption{Computational cost Batch Verification}
\label{fig6}
\end{figure}

In the proposed QS-CL-CPPA scheme, we refer three phases $CL_{sig}-Gen()$, $CL_{sig}-Ver()$, and $Batch-CL_{sig}-Ver$ for discussion of computational costs analysis. In the phase $CL_{sig}-Gen()$, the proposed QS-CL-CPPA scheme consumes 49.473 ms., while the state-of-the-art schemes  \cite{Li2022lattice, mukherjee2019an, Liu2019a}, and \cite{Dharminder2020LCPPA} consume 447.391 ms., 49.785 ms., 50.184 ms., and 49.785 ms. respectively (Table \ref{tab4}). Thus, the $CL_{sig}-Gen()$ phase of the proposed QS-CL-CPPA scheme is better (Fig. \ref{fig4}) than the compared ones. Single signature verification has been considered in the phase $CL_{sig}-Ver()$. The QS-CL-CPPA scheme takes 49.467 ms. and related literature \cite{Li2022lattice, mukherjee2019an, Liu2019a}, and \cite{Dharminder2020LCPPA} consume 60.118 ms., 49.779 ms., 49.785 ms., and 49.779 ms. respectively (Table \ref{tab4}). Thus, the scheme in QS-CL-CPPA consumes the least cost (Fig. \ref{fig5}), i.e. phase $CL_{sig}-Ver()$ is also better than the related schemes. For the phase $Batch-CL_{sig}-Ver$, the batch sizes 50, 100, 150, 200, 250, and 300 have been taken. As per Fig. \ref{fig6}, the proposed QS-CL-CPPA scheme takes the least computational cost in this phase. 
\begin{table*}[h!]
\caption{Comparison of  Computational Cost} \label{tab4}\centering
\begin{tabular}{|l|l|c|c|} 
\hline   \textbf{ Schemes }  & $CL_{sig}-Gen()$ & $CL_{sig}-Ver()$ & $Batch-CL_{sig}-Ver$  \\
\hline Li {\it et al.} \cite{Li2022lattice}  & $\begin{array}{l}
E_{{h }}+\\ 
E_{{a d d}-{v_m}}+. \\
E_{\text {{m u l }-{$v_q$ y A}}}+ \\
E_{mul-mat} \approx \\
447.391 {~ms}\end{array}$ & $\begin{array}{l}E_{\text {m u l-$v_q$ A }}+   \\
2 E_{\text {{m u l}-{$v_q$ x A} }}+   \\
2 E_{\text {{a d d}-{$v_n$} }} \approx \\
60.118 m s\end{array}$ & $\begin{array}{l}2 N E_{\text {{m u l}-{$v_q$ x A} }}+ \\
3 N  E_{\text {{m u l }-{$v_q$ d y} }}+ \\
N  E_{\text {{m u l }-{$v_q$ c x} }}+ \\
E_{\text {m u l-$v_q$ A }}+(3 N+ 2) \\ E_{\text {{a d d}-{$v_n$} }}+ \\
N  E_{\text {a d d}-{v_m}} \approx \\
48.815 + 12.784N {~ms}\end{array}$  \\
\hline Mukherjee {\it et al.} \cite{mukherjee2019an}   & $\begin{array}{l}\\
 E_{\text {m u l-$v_q$ A }}+ \\
 E_{ {h }}+
2 E_{\text {{m u l }-{$v_q$ c x}}}+ \\
 E_{\text {a d d}-{v_m}} \approx \\
49.785 {~ms}\end{array}$ & $\begin{array}{ll}E_{\text {m u l-$v_q$ A }}+ &  \\
2 E_{\text {{m u l }-{$v_q$ c x} }}+ &  \\
E_{\text {a d d - $v_m$ }} & \approx \\
49.779 {~ms} & \end{array}$ & $\begin{array}{l}E_{{m u l}-{v_q A}}+2(3 N+ 1) \\
\text E_{{m u l }-{v_q c x}} + \\
N E_{a d d- v_m} \approx\\
49.394 + 4.57 N {~ms}\end{array} $ \\

\hline Liu {\it et al.} \cite{ Liu2019a}   & $\begin{array}{l}\\
 E_{\text {m u l-$v_q$ A }}+ \\
3 E_{{h }}+ E_{ {x o r}} +\\
\text2 E_{{m u l }-{v_q c x}} + \\
 +2 E_{\text {a d d}-{v_m}} \approx \\
50.184 {~ms}\end{array}$ & $\begin{array}{ll}E_{\text {m u l-$v_q$ A }} +& \\
2E_{\text {{m u l }-{$v_q$ c x}}}+ \\
E_{{h }}+ &  \\
E_{\text {a d d - $v_m$ }} & \approx \\
49.785 {~ms} & \end{array}$ & $\begin{array}{l}E_{{m u l}-{v_q A}}+2 (3N+1)\\
E_{\text {{m u l }-{$v_q$ c x}}}+ \\
2N E_{a d d- v_m} \approx\\
49.394+ 4.952 N{~ms}\end{array} $  \\

\hline Dharminder {\it et al.} \cite{Dharminder2020LCPPA}  & $\begin{array}{l} 
E_{\text {m u l-$v_q$ A }} + E_{ {h }}+ \\
2 E_{\text {{m u l }-{$v_q$ c x}}}+ \\
 E_{\text {a d d}-{v_m}} \approx \\
49.785 {~ms}\end{array}$ & $\begin{array}{ll}E_{\text {m u l-$v_q$ A }} +& \\
2 E_{\text {{m u l }-{$v_q$ c x} }}+ &  \\
 E_{\text {m u l-$v_m$ }} & \approx \\
49.779 {~ms} & \end{array}$ & $\begin{array}{l}E_{{m u l}-{v_q A}}+ 2(3 N+ 1) \\
\text E_{{m u l }-{v_q c x}} + \\
 N 
E_{a d d- v_m} \approx\\
48.394 + 4.567N {~ms}\end{array} $ \\

\hline  $\begin{array}{l}\text {QS-CL-CPPA } \\
\text {scheme }\end{array}$   & $\begin{array}{l}
E_{\text {m u l-$v_q$ A }}+ \
E_{\text {h }}+ \\
E_{\text {m u l-$v_q$ c x }}+ \\
2 E_{\text {a d d }-{v_m}} \approx \\
49.473 {~ms}\end{array}$ & $\begin{array}{ll}E_{\text {m u l-$v_q$ A }} & + \\
 E_{\text {{m u l}-{$v_q$ c x}}}+ & \\
2 E_{\text {a d d}-{v_m}} & \approx \\
49.467 {~ms} & \end{array}$ & $\begin{array}{l}E_{\text {m u l-$v_q$ A }}+(3 N+ 1) \\
E_{\text {{m u l }-{$v_q$ c x} }} + \\
2 N  E_{ {a d d}-{v_m}} \approx  \\
48.697 + 2.681 N {~ms}\end{array} $  \\
\hline

\end{tabular}
\end{table*}

\subsection{Communication Cost Analyses}

\begin{table}[ht]
\caption{Comparison of Communication Cost} \label{tab5}\centering
\begin{tabular}{|l|l|l|}
\hline Schemes & $\begin{array}{l}\text {Msg-Sig} \\
\text {Pair}\end{array}$ & $\begin{array}{l}\text {N Msg-Sig} \\
\text {Pairs}\end{array}$ \\
\hline Li {\it et al.} \cite{Li2022lattice} & 7757 & $7757^* {~N}$ \\
\hline Mukherjee {\it et al.} \cite{mukherjee2019an}  &850 & $850^* {~N}$ \\
\hline Liu {\it et al.} \cite{Liu2019a} & 850& $850^* {~N}$\\

\hline Dharminder {\it et al.} \cite{Dharminder2020LCPPA} & 850 & $850^* {~N}$ \\

\hline QS-CL-CPPA & 762 & $762^* {~N}$ \\
\hline
\end{tabular} \end{table} 
To discuss the communication cost analysis, we consider the same notation of anonymous ID as $ANS_i$, given in most related literature. In the CPPA protocol of Li {\it et al.} \cite{Li2022lattice}, the final signature is sent by a user to other nearby nodes $ (A N S_i, T_i, {Z}_i, sig_i)$. Since $A N S_i= ({A S N}_{i, 1},{A N S}_{i, 2})$ and ${A N S}_{i, 2}$  $= R I D \oplus H_1({S}_{{i}}) \bmod q, {A N S}_{i, 2}$ are random elements in ${Z}_q^n$, authors took  all values of ${A N S}_{i, 2}$ be $q-1$ to maximize the size of ${A N S}_{i, 2}$. likewise, ${A N S}_{i, 1}$ is chosen randomly from ${Z}_q^n$, the maximum size of $A N S_i$, which is $2 n * \log _2(q-1)$ = 1400 bits. Also the maximum size of ${Z}_i$ is $n * k * \log _2(q-1)$ = 56000 bits, for vector is $q-1$  and for $\operatorname{sig}_i$ is $m * \log _2(q-1)$ = 4662 bits. Therefore, the maximum size of the total signature $\left(A N S_i, T_i, \mathbf{Z}_i,
\operatorname{sig}_i\right)$ is approx 7757 bytes.

\par In the Mukherjee {\it et al.}'s CPPA scheme \cite{mukherjee2019an}, the maximum size of $A N S_i$ is $2 * n * \log _2(q-1)$ = 1400 bits and  that of $\mathbf{R}_i$ is $n * \log _2(q-1)$ = 700 bits and for $\sigma_i$, the maximum size is $m * \log _2(q-1)$ = 4662 bits. Therefore, the signature ( $\left.M_i, A N S_i, T_i, \mathbf{R}_i, \sigma_i\right)$  has a maximum size of approximately 846 + 4 = 850 bytes.

\par In the Liu {\it et al.} scheme \cite{Liu2019a}, the maximum size of $A N S_i$ is $2 * n * \log _2(q-1)$ = 1400 bits and  that of $\mathbf{R}_i$ is $n * \log _2(q-1)$ = 700 bits and the size of $\sigma_i$ is $m * \log _2(q-1)$ =4662 bits. Therefore, the overall signature ( $\left.M_i, A N S_i, T_i, \mathbf{R}_i, \sigma_i\right)$  has a maximum size of approximately 846 + 4 = 850 bytes.
 
\par In the CPPA protocol of Dharminder {\it et al.} \cite{Dharminder2020LCPPA}, the maximum size of $A N S_i$ is $2 * n * \log _2(q-1)$ = 1400 bits and  that of $\mathbf{R}_i$ is $n * \log _2(q-1)$ = 700 bits. Also, the maximum size of $\sigma_i$ is $m * \log _2(q-1)$ = 4662 bits. Therefore, the final signature ( $\left.A i d_i, T_i, \mathbf{R}_i, \sigma_i\right)$  has a maximum size of approx 846 + 4, which equals to 850 bytes.
\par In the QS-CL-CPPA scheme, the final signature sent by one node to the other nodes is $(P_{{ID}_i}, T_i, \mathbf{R}_i, \sigma_i)$. The maximum size of $P_{{ID}_i}$ was calculated as $ n * \log _2(q-1)$ = 700 bits and  total size of $\mathbf{R}_i$ is $n * \log _2(q-1)$ = 700 bits. Additionally, the maximum size for $\sigma_i$ turned out $m * \log _2(q-1)$ = 4662 bits. Therefore, in the $QS-CL-CPPA$ protocol, the overall signature $(P_{{ID}_i}, T_i, \mathbf{R}_i, \sigma_i)$  has a maximum size of approximately 757.75 + 4 = 761.75 bytes.
\par The comparison result of communication costs of the above four schemes \cite{Li2022lattice}, \cite{mukherjee2019an}, \cite{Liu2019a}, \cite{Dharminder2020LCPPA} with this scheme is displayed in Table \ref{tab5}. Overall, this comparison shows that our  QS-CL-CPPA scheme has less communication cost than the compared ones.
\par Thus, as per the above analysis, the proposed QS-CL-CPPA scheme is better in terms of both computational cost and communication cost than the state-of-the-art \cite{Li2022lattice}, \cite{mukherjee2019an}, \cite{Liu2019a}, \cite{Dharminder2020LCPPA}. Besides, according to Table \ref{tab1}, the proposed QS-CL-CPPA scheme has better security as the secret key of the vehicle is not generated by TA, and the secret key of TA is not given to the vehicle. Therefore, the proposed QS-CL-CPPA scheme is the most suitable design to deploy for post-quantum scenarios on VANETs.  

\section{Conclusion}
To improve safety in traffic systems, the paradigm of VANETs has been introduced by the amalgamation of communication technologies and automobile technologies. However, wireless channels in VANETs are prone to adversarial attacks. This paper introduced a novel and efficient QS-CL-CPPA scheme to diminish such attacks. The devised QS-CL-CPPA scheme addresses the potential flaws of master secret-key leakage and key-escrow problems in the existing literature. Relying on the hardness of SIS/ISIS problem, the security analysis validates its resistance to forgery attacks. Employing the batch signature technique leads to notable efficiency enhancements in the proposed QS-CL-CPPA. A comprehensive comparison of the proposed QS-CL-CPPA scheme to the state-of-the-art schemes discloses that this scheme can eliminate the drawbacks of a post-quantum scenario. Lastly, a rigorous analysis of performance analysis assessment of the scheme reflects its suitability for deployment in VANETs.
\par The proposed QS-CL-CPPA scheme needs a secure channel to establish the key of a user. Thus, as a future work, this limitation should be eliminated.

\end{document}